\documentclass[]{elsart}
\usepackage{epsfig}
\usepackage{times}
\usepackage{helvet}
\usepackage{latexsym}
\begin{document}
\begin{frontmatter}

%
    
    \title{ Centrality Dependence of the High $p_T$ Charged Hadron 
    Suppression  in  Au+Au Collisions at $\sqrt{s_{NN}}=130$ GeV}  

%

\author{K.~Adcox,$^{40}$}
\author{S.{\,}S.~Adler,$^{3}$}
\author{N.{\,}N.~Ajitanand,$^{27}$}
\author{Y.~Akiba,$^{14}$}
\author{J.~Alexander,$^{27}$}
\author{L.~Aphecetche,$^{34}$}
\author{Y.~Arai,$^{14}$}
\author{S.{\,}H.~Aronson,$^{3}$}
\author{R.~Averbeck,$^{28}$}
\author{T.{\,}C.~Awes,$^{29}$}
% ...
    \author{(PHENIX Collaboration)}
    \address{
% ...
$^{3}$Brookhaven National Laboratory, Upton, NY 11973-5000, USA\\
% ...
$^{14}$KEK, High Energy Accelerator Research Organization, Tsukuba-shi, Ibaraki-ken 305-0801, Japan\\
% ...
$^{27}$Chemistry Department, State University of New York - Stony Brook, Stony Brook, NY 11794, USA\\
    $^{28}$Department of Physics and Astronomy, State University of New York - Stony Brook, Stony Brook, NY 11794, USA\\
    $^{29}$Oak Ridge National Laboratory, Oak Ridge, TN 37831, USA\\
% ...
$^{34}$SUBATECH (Ecole des Mines de Nantes, IN2P3/CNRS, Universite de Nantes) BP 20722 - 44307, Nantes-cedex 3, France\\
% ...
$^{40}$Vanderbilt University, Nashville, TN 37235, USA\\
% ...
}

\date{\today}        

%
\begin{abstract}
    PHENIX has measured the centrality dependence of charged hadron
    $p_T$ spectra from Au+Au collisions at
    $\sqrt(s_{NN})=130$~GeV. The truncated mean $p_{T}$ decreases with 
    centrality for $p_T >2$~GeV/c, indicating an apparent reduction of
    the contribution from hard scattering to high $p_T$ hadron
    production. For central collisions the yield at high $p_T$ is
    shown to be suppressed compared to binary nucleon-nucleon
    collision scaling of p+p
    data. This suppression is monotonically increasing with centrality,
    but most of the change occurs below 30\% centrality, i.e. for
    collisions with less than $\sim$140 participating nucleons. The 
    observed $p_T$ and centrality
    dependence is consistent with the particle production predicted
    by models including hard scattering and subsequent energy
    loss of the scattered partons in the dense matter created in the
    collisions.

\end{abstract}
\end{frontmatter}

%
%
\section{Introduction}
Particle production at large transverse momentum ($p_T$) provides a
new tool to study hot and dense nuclear matter created in
high energy nuclear collisions. In nucleon-nucleon collisions,
hadrons with $p_T \geq 2$~GeV/c are believed to originate mostly 
from the jet fragmentation of constituent partons, quarks and gluons, that 
were scattered with large momentum transfer $Q^2$ \cite{E605}. 
In nuclear collision these hard scattering processes between
constituent partons occur early compared to the lifetime of the 
strongly interacting matter. Thus the hard scattered partons may
traverse the highest energy density matter produced. Theoretical
studies of the propagation of partons in high density matter suggest 
that they lose a significant fraction of their energy through gluon 
bremsstrahlung \cite{jet-quenching} and that the energy lost reflects 
the density of color charges in the matter through which they pass
\cite{quench-density-dep}. The energy loss reduces the momenta
of the partons, which results in a corresponding reduction of the
momenta of the fragmentation products \cite{quench-effect},
observable as a reduced yield of high $p_{T}$ hadrons. 

The first measurements of hadron spectra at the Brookhaven National
Laboratory Relativistic Heavy Ion Collider (RHIC) facility
indicate a suppression of high-$p_T$ hadron production in central
Au+Au collisions relative to a binary collision scaling of p+p and
$\bar{\mathrm p}$~+p data \cite{ppg003,star}. No suppression is 
found for peripheral Au+Au collisions. So far no 
unique explanation of this apparent absence of the expected jet 
contribution to the $p_T$ spectrum above 2 GeV/c has been identified. 

Models of parton energy loss can reproduce the observed suppression 
in central Au+Au collisions \cite{wang-2,GLV,GLV-discovery}. 
Other final state effects such as re-scattering of hadrons originally 
produced via the jet fragmentation have been proposed to explain 
the suppression \cite{gallmeister}. It should be noted that models 
invoking thermal hadron
production combined with collective transverse expansion of the
reaction volume successfully describe the transverse momentum 
distributions of identified hadrons up to 3 GeV/c \cite{kolb,teaney}. 
However, the mechanism of equilibration, which requires a reduction of
the high $p_T$ particle yield, is not specified in these models.

Alternatively, the initial state may be modified such that the number
of hard scatterings is reduced. It is well known that nuclear
modifications of the parton distributions exist \cite{EMC}. These
modifications cannot explain the suppression, since in the kinematic
range of the measurements anti-shadowing enhances the parton
distributions in nuclei \cite{Gribov,sarcevic,eskola}. However, models
using a classical QCD picture of a highly relativistic nucleus 
\cite{mclerran,kovner} suggest that gluon distributions are saturated
for momenta below a scale $Q_{s}$ and thus reduced compared to
expectations based on perturbative QCD \cite{krasnitz}. As a
consequence, a considerable suppression of hadron production might be
expected even well above $Q_{s}$ \cite{mclerran-2}.


In this letter we present the centrality dependence of the suppression
of the high-$p_T$ hadron yield to provide new experimental constraints
on theoretical descriptions. These data are complementary to the
previous study of the absolute yields \cite{ppg003} and have different 
systematic errors.  

\section{Experimental setup and data analysis}

%
The results are obtained from $1.4 \times 10^6$ mini\-mum bias Au+Au 
col\-lisions at $\sqrt{s_{NN}}$=130 GeV recorded by the PHENIX
experiment during the Run-1 operation of RHIC 
(August - September 2000). Details on the PHENIX detector and 
its configuration in Run-1 operation can be found in \cite{ppg003,ppg002}.

In PHENIX semi-inclusive charged hadron spectra are measured over 
the range $0.5 < p_T < 5.0$~GeV/c in the east central arm spectrometer
using data from a drift chamber (DC) and two segmented cathode pad 
chambers (PC1 and PC3), located outside of an axial magnetic field 
at a radial distance of 2.2, 2.5 and 5~m from the beam axis. The 
detectors cover an azimuthal 
acceptance of $90^\circ$ and a pseudo-rapidity range of $| \eta |
<0.35$. In this analysis an additional fiducial cut 
$|\eta| <0.18$ is applied to guarantee homogeneous track acceptance for
collisions within $|z_{vtx}| < 30$~cm of the nominal interaction
point. About $\sim$25\% of the azimuthal acceptance is covered
by a time-of-flight system which allows proton identification out to
3.5 GeV/c, where the measurement is limited by statistics \cite{ppg006}. 

A pair of beam-beam counters provides the vertex position 
along the beam direction (z). Each charged track is reconstructed 
from the DC measurements of its projection into the bend plane of 
the magnetic field and two space points provided by 
PC1 and PC3. The unphysical background, resulting from 
false associations of drift chamber projections with pad chamber
points, is estimated and subtracted by forming artificial events with
the locations of pad chamber points inverted around the symmetry axis
of the spectrometer. Physical background from decays in flight and 
photon conversions close to the DC, which only partially traverse the
field and thus mimic high momentum tracks,
are removed by requiring the track to point back to the event vertex 
within $|z_{vtx}|< 2.5$~cm. The remaining background level 
is negligible below 4~GeV/c and less than 40\% at 5~GeV/c; this upper 
estimate is included in the systematic errors. 

\begin{figure}
 \begin{center}
     \epsfig{file=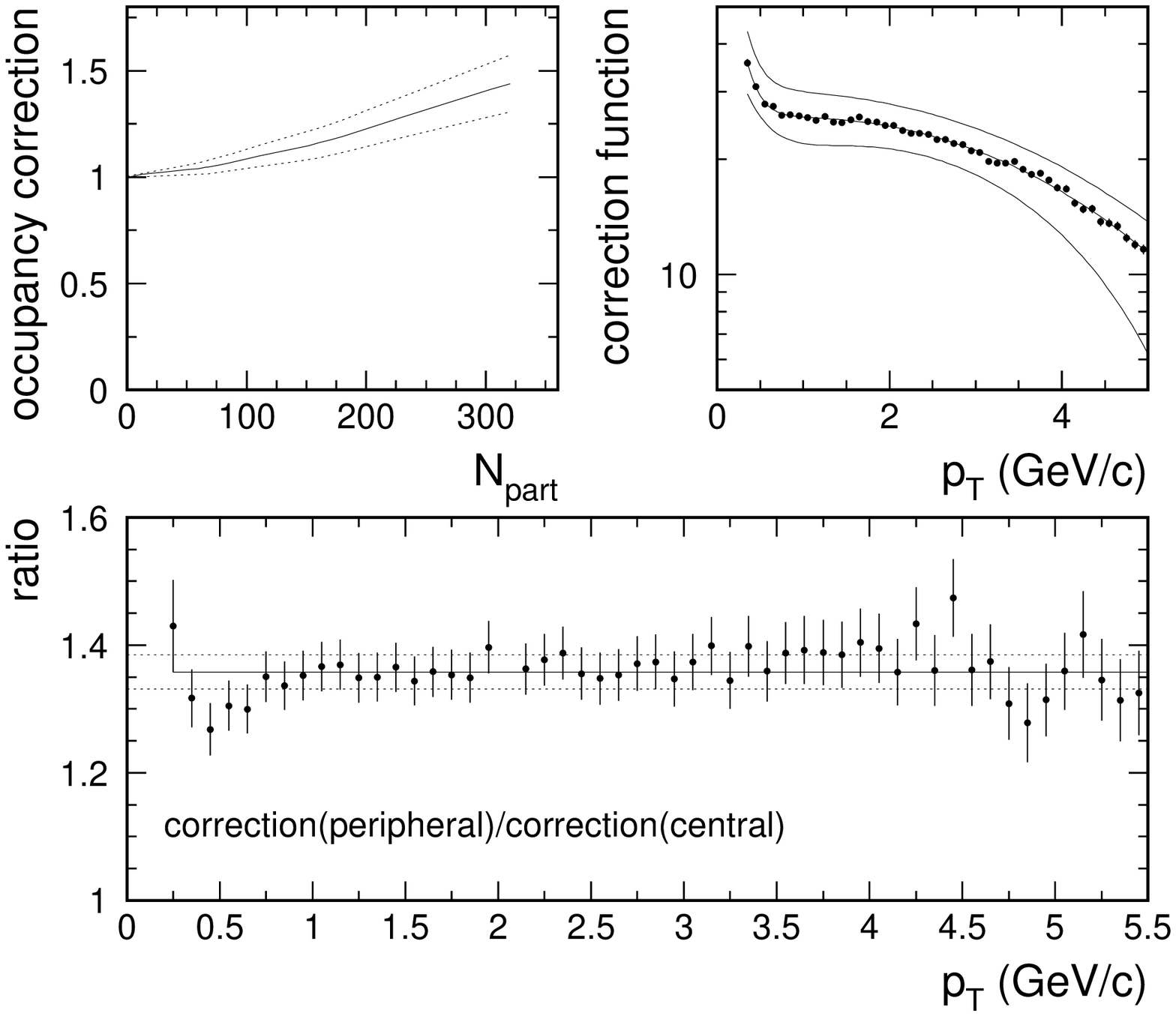,width=1.0\linewidth}
     \caption{\label{fig:corr}
	Functions used to correct the charged particle $p_T$ spectra. Upper 
	left panel shows the centrality dependent correction
        $c(N_{part})$ and the right panel shows the $p_T$ dependent 
	correction $c(p_T)$. The systematic uncertainties are
        indicated by the dashed lines. The two corrections factorize, so
        that for any centrality the full correction function is given by 
	$c(p_T) \times c(N_{part})$. The accuracy of this factorization 
	is demonstrated in the lower panel. The ratio of the full
        correction for central collisions (top 5\%) to the correction
        for single particle events 
	varies by less than $2$\% above 1 GeV/c. 
	} 
\end{center}
\end{figure}

Corrections of the data are determined by tracing individual particles 
through a full GEANT simulation, simulating the detector response,
then merging this response with that of all particles from a real
event and passing the composite event through the PHENIX reconstruction
software.  
The average track reconstruction efficiency in the active detector
area is larger than 98\% in peripheral collisions and decreases 
to 68$\pm$6\% for central collisions. The corresponding correction 
is shown on the upper left hand side of  Fig.~\ref{fig:corr}. The full 
correction also depends on $p_T$. It is plotted for peripheral collision on 
the upper right hand side of the figure. Between 0.8 and 2.5~GeV/c
the correction factor varies slowly with $p_T$. Its value of $\sim 25$ 
corrects for geometrical acceptance ($\Delta\phi = \pi/4$; 
$\Delta\eta = 0.36$), dead areas of the detectors ( 45\% DC, 5\%
PC's), and losses due to $2\sigma$ track matching cuts. At lower $p_T$ 
the correction increases reflecting the gradual loss of acceptance.
At higher $p_T$ the observed particle 
yield is artificially increased because of the 
finite momentum resolution ($\delta p/p \simeq 0.6\%  \oplus 3.6\% \
p \,(\mathrm{GeV}/c)$) and therefore the correction function decreases. 
Since this correction depends on the spectral shape of the true $p_T$
distribution it was determined iteratively. At 5 GeV/c the correction 
is reduced by a factor of $\sim$2. The
systematic uncertainties are indicate by the dashed lines; they are
also tabulated in Table~\ref{tab:sys}. As shown in the lower part of
Fig.~\ref{fig:corr} correction factorizes into functions of centrality
(i.e. detector occupancy) and $p_T$ within 2\% systematic uncertainty 
in the range from 2 to 5 GeV/c. 

%
Events are selected according to centrality following the procedure
described in \cite{ppg003}. Six exclusive centrality bins are 
established using the energy measured in two zero-degree calorimeters 
and the number of charged-particles detected in the two beam-beam
counters. A Monte-Carlo simulation using measured nucleon density 
distributions calculated in the Glauber eikonal approximation was used 
to estimate the average number of binary nucleon-nucleon collisions 
($N_{coll}$) and the corresponding average number of participating 
nucleons ($N_{part}$) for each bin. The results are quoted in
Table~\ref{tab:1}. 
%
%
\section{Results}

\begin{figure}
 \begin{center}
     \epsfig{file=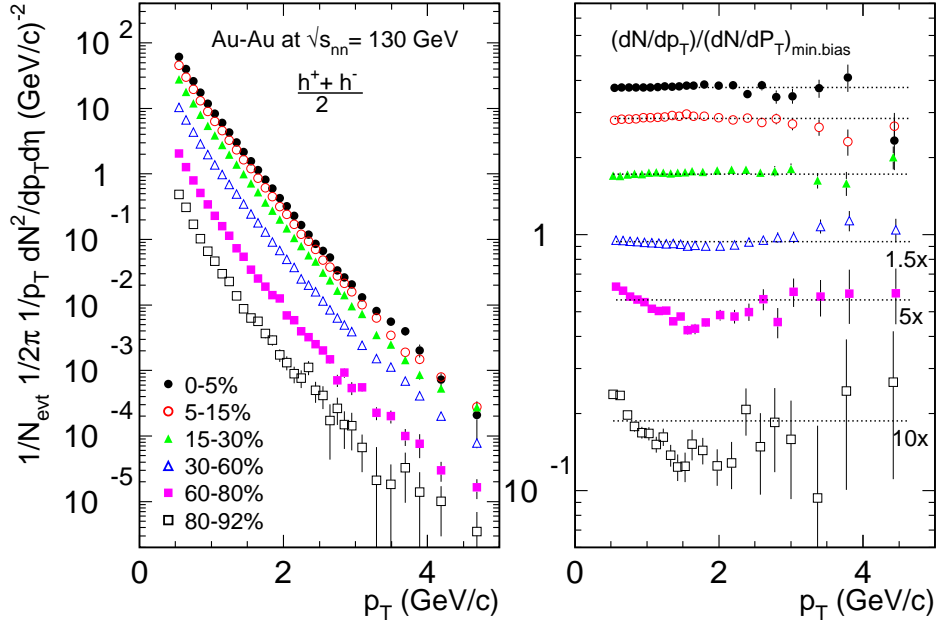,width=1.0\linewidth}
     \caption{\label{fig:spectra}
	The left panel shows $p_T$ spectra of charged hadrons from six
	Au+Au centrality selections. Error bars indicate statistical errors
	only. The $p_T$ dependent systematic errors are  independent
        of centrality and not shown, they are given in
        Table~\ref{tab:sys}. The centrality dependent errors are less than
        10\% and small compared to the symbol size. The right panel
     	shows the ratio of each of the centrality selected $p_T$
	spectra to the minimum bias spectrum. Ratios for
	peripheral selections are scaled for clarity. Dotted lines indicate
	the average ratios for each centrality selection.
	} 
\end{center}
\end{figure}

Fig.~\ref{fig:spectra} presents charged hadron $p_T$ spectra 
for the six centrality bins. For peripheral collisions the spectra   
are more concave than those for central collisions. This shape
difference is seen more clearly by taking the ratio of the spectrum
for each centrality bin to the minimum-bias spectrum, as shown on the
right hand side of Fig.~\ref{fig:spectra}. In these ratios most 
systematic errors cancel or affect the overall scale only. 
The ratios for the central bins are almost independent of $p_T$ since
central collisions dominate the minimum bias yields. 
The peripheral bins show a decreasing ratio between 0.5 and 1.5~GeV/c 
thus in peripheral collisions the yield in this region decreases more 
rapidly with increasing $p_T$ than in central collisions. 
For $p_T$ above 1.5 GeV/c this trend is inverted.

\begin{figure}
  \begin{center}
      \epsfig{file=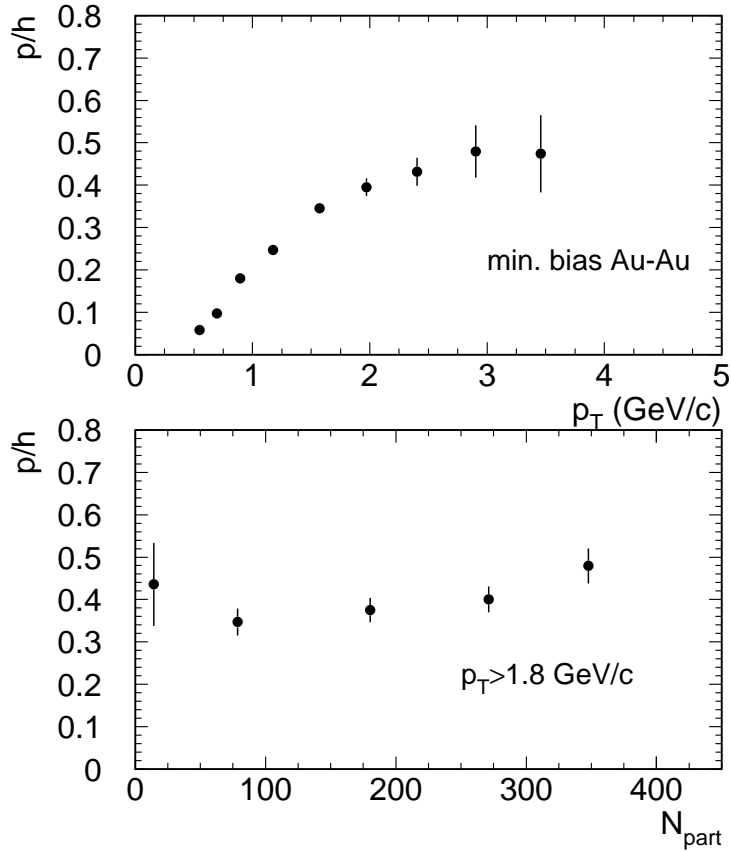,width=0.8\linewidth}
      \caption{\label{fig:h2p}
	The ratio $p/h$ represents the proton plus anti-proton yield 
	relative to the total charged hadron multiplicity. The top panel
	shows the $p_T$ dependence of $p/h$ for minimum bias events. 
	In the bottom panel we show the centrality dependence of $p/h$ 
	for $p_T > 1.8$ GeV/c. Only statistical errors are shown.  
	}
  \end{center}
\end{figure}

%
%
Before analyzing the centrality dependence in more detail we demonstrate
that the observed suppression of the yield at high $p_T$ does not 
result from a reduced yield of protons and anti-protons \cite{ppg006}. 
To evaluate the effect of the (anti-)protons, we plot in the top panel 
of Fig.~\ref{fig:h2p} the $p_T$ dependence of $p/h$, the ratio of
proton plus anti-proton yields to the total charged hadron yield for 
minimum bias collisions, which increases steadily. 
Above 1.5~GeV/c the ratio seem to saturate reaching a value of $\sim
0.5$ arround 3 GeV/c. In the bottom panel of Fig.~\ref{fig:h2p} we show 
$p/h$ for $p_T$ above 1.8 GeV/c as a function of centrality.  Since 
there is clearly no significant decrease of the $p/h$ ratio 
with either centrality or $p_T$, the observed hadron suppression is not
due to a larger suppression of the (anti-)proton component than that
of the mesons. On the contrary, within the range of the presented 
measurement the apparent slight increase in $p/h$ with $N_{part}$ could indicate
a larger suppression of the meson component relative to all charged
hadrons.

\begin{figure}
  \begin{center}
      \epsfig{file=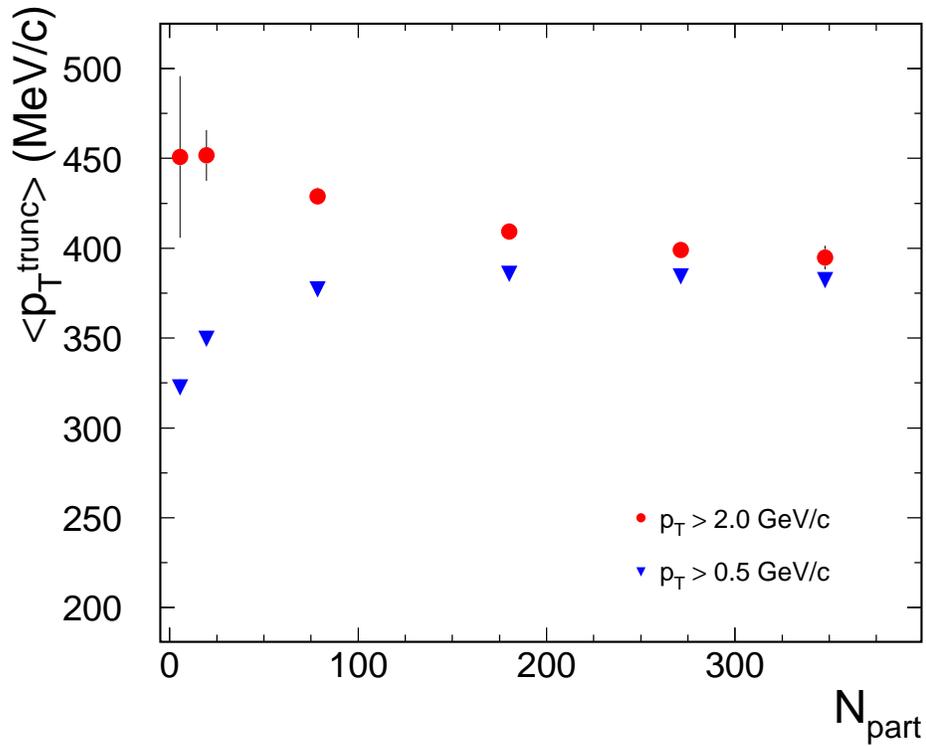,width=1.0\linewidth}
      \caption{\label{fig:avg}
	Centrality dependence of $\langle p_T^{trunc} \rangle$, the 
	average $p_T$ of charged particles with $p_T$ above a threshold
	$p_T^{min}$ minus the threshold $p_T^{min}$. Shown are values for 
	two $p_T^{min}$ cuts, one at $p_T >$~0.5 GeV/c representing all 
	data presented in Fig.~\ref{fig:spectra} and the other one at 
	$p_T >$~2~GeV/c. Only statistical errors are shown; see the text
        following Eq.~\ref{eq:x} for a discussion of the sytematic errors.
	}
  \end{center}
\end{figure}

%
%
To evaluate the change of the hadron spectra more quantitatively we 
calculate the truncated average $p_T$:
\begin{equation}
     \langle p_T^{trunc} \rangle \equiv
     \frac{\int_{p_{T}^{min}}^{\infty} p_{T} \cdot dN/dp_{T}}
     {\int_{p_{T}^{min}}^{\infty} dN/dp_{T}} - p_{T}^{min}
     \label{eq:x}
\end{equation}
for $p_T^{min} = 0.5$~GeV/c and for $p_T^{min} = 2.0$~GeV/c
for each centrality selection 
\footnote{The value of $\langle p_T^{trunc} \rangle $ is closely
related to the local inverse slope which is slightly smaller. The
conversion to local slope depends on the spectral shape and also on
$p_T^{min}$. For an exponential spectrum with an inverse slope of 
350~MeV/c the conversion is approximately 
$-80 (-60)$~MeV/c for $p_T^{min} = 0.5 (2.0)$~GeV/c.}.
In Fig.~\ref{fig:avg} $\langle p_T^{trunc} \rangle$ is plotted as 
a function of $N_{part}$. The value of $\langle p_T^{trunc}
\rangle$ is insensitive to the normalization of the spectra.
Systematic uncertainties of $\sim$20(50) MeV/c for  
$p_T^{min} = 0.5(2.0)$~GeV/c result from an 1-2\% uncertainty on 
the momentum scale and the $p_T$ dependent uncertainties on the
correction function. Since the centrality and $p_T$ dependence of the
correction factorize the error on  $\langle p_T^{trunc} \rangle$ is 
independent of centrality to better than 2 MeV/c. 

The $\langle p_T^{trunc}\rangle $ for $p_T^{min} = 0.5$~GeV/c
increases with $N_{part}$ similar to the average $p_T$ of identified 
charged hadrons \cite{ppg006}. For higher $p_T^{min} = 2.0$~GeV/c
the $\langle p_T^{trunc}\rangle$ drops by $\sim$60~MeV/c with
$N_{part}$, distinctly different from the expected increased role  
of hard particle production in the more central collisions. Absent any
collective effects, the hard scattering
contribution should increase relative to soft production by the factor 
$N_{coll}/N_{part}$ which grows from 1.5 to $\sim 6$ from peripheral 
to central collisions. Since the relative contribution of the hard component to the
spectrum increases with $p_T$, this should lead to a rise of 
$\langle p_T^{trunc} \rangle$ for sufficiently large $p_T^{min}$. 
The drop of $\langle p_T \rangle$ therefore indicates the suppression 
of the high $p_T$ relative to the low $p_T$ hadron yield  
independent of systematic errors associated with the absolute 
normalization of the spectra or any nucleon-nucleon 
reference distribution. 

%
%
Changes of the hadron spectra at high $p_T$ are often presented  
in terms of the nuclear modification factor $R_{AA}$. This measure  
relies on the absolute normalization and a reference and therefore has
intrinsically larger systematic uncertainties, but it allows to 
quantify the suppression. We have calculated $R_{AA}$ for each 
centrality bin as:  
\begin{equation}
R_{AA}(p_T,\eta) 
=  (\frac{1}{N_{evt}} \frac{d^{2}N^{A+A}}{dp_T d\eta})  /
(\frac{\langle N_{coll} \rangle}{\sigma^{N+N}_{inel}}
\frac{d^{2}\sigma^{N+N}}{dp_T d\eta}) 
\label{eq:RAA_defined}
\end{equation}

\begin{figure}
  \begin{center}
      \epsfig{file=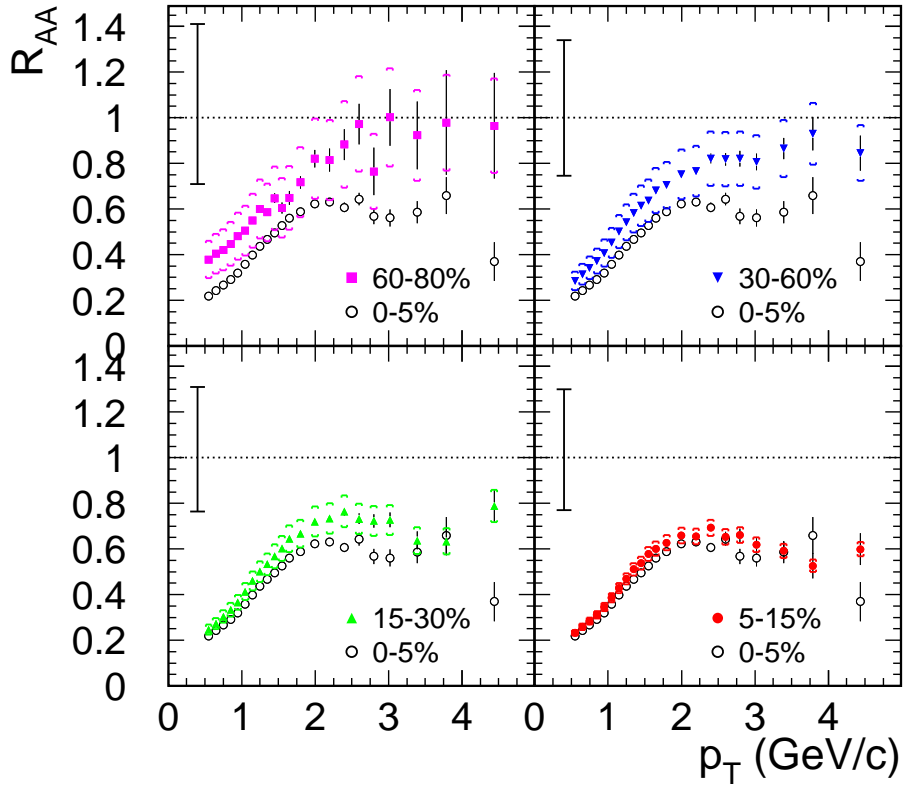,width=1.0\linewidth}
      \caption{\label{fig:raa} 
	Nuclear modification factor ($R_{AA}$)
	for the 60-80\%, 30-60\%, 15-30\%, and 
	5-15\% centrality selections compared to the one for the most 
	central sample (0-5\%). Due to insufficient statistics $R_{AA}$ is 
	not shown for the 80-92\% sample. The solid error bars on each 
	data point are statistical. The systematic error between the more 
	peripheral and the central sample are given as brackets for the 
	more peripheral data points. The error bar on the left hand side 
	of each panel indicates the overall systematic error on the 
	$R_{AA}$ scale. 
	}
  \end{center}
\end{figure}

For the N+N charged hadron cross section we use a power-law
parameterization $1/\pi \  d^2\sigma/dp_T^2 = A/(1+ p_T/p_0)^n$, 
with $A=$330~mb/(GeV/c)$^2$, $p_{0}= 1.72 \, \mathrm{GeV}/c$, 
and $n=12.4$. The parameters were obtained by interpolating p+p 
and $\bar{\mathrm p}$+p data to $\sqrt{s} = 130$~GeV as described 
in \cite{ppg003}.  In Fig.~\ref{fig:raa} the $R_{AA}(p_T)$ 
values for all centrality bins excluding the most peripheral one 
are compared to the central (0-5\%) bin. The systematic uncertainties in 
the normalization of the data, in $N_{coll}$, and in the 
N+N reference (20\%) result in overall systematic errors
of about 41\%, 34\%, 31\%, 31\% and 30\% for centrality bins 2-6, 
respectively. The errors are quoted for the range from 1 to 3.5~GeV/c;
they increase somewhat towards higher $p_T$. Comparing $R_{AA}$ for 
a given centrality bin to the most central bin, the systematic errors reduce 
to $\sim 6\%$ for bin 5 (5-15\%) and nearly 27\% for bin 2(60-80\%). 
They are dominated the uncertainty in $N_{coll}$ listed in 
Table~\ref{tab:1}. 

For the 60-80\% centrality bin, $R_{AA}$ increases with $p_T$ and reaches 
unity at high $p_T$. In comparison to the 60-80\% bin, the $R_{AA}$ 
values for the most central bin remain significantly below unity at 
a value of 0.55 for $p_T > 2$~GeV/c. At high $p_T$ approximately
constant $R_{AA}$ values are detected in all centrality bins. 
The high $p_T$ values decrease monotonically with centrality, 
clearly indicating that the 
magnitude of the suppression of high $p_T$ hadrons increases with 
centrality. This is shown clearly in the upper part of
Fig.~\ref{fig:yields}, which presents $R_{AA}$ obtained for the three 
$p_T$ bins 1.6 to 2.6~GeV/c, 2.6 to 3.6 GeV/c, and above $3.6$~GeV/c 
as a function of centrality. For central collisions we observe a 
suppression of about a factor of $2\pm 0.6$ compared to binary collision
scaling. Relative to peripheral collisions the suppression factor in central 
collisions increases with $p_T$ from $1.25\pm0.2$ to $1.5\pm0.2$ 
to $1.8\pm0.3$ for the three $p_T$ bins, respectively. 

We note that for peripheral collisions the data do not indicate a 
significant increase of $R_{AA}$ above unity, unlike data at lower 
energies \cite{wang-enke}. However, such an increase, attributed to 
initial state scattering, the Cronin effect \cite{cronin}, may well be 
consistent with the peripheral data due to the large systematic uncertainty 
of the $R_{AA}$ scale. While the relative difference between the 
peripheral and central spectra increases with $p_T$, the roughly
constant nuclear modification factor at large $p_T$ suggests an approximately 
$p_T$ independent suppression of hard scattering contributions over the
range $2 < p_T < 4.5$~GeV/c. 

%
%
The physics that controls the production of high-$p_T$
particles or the suppression of the hard scattering yields in the
measured $p_T$ range may not depend directly on $N_{coll}$. Thus, we 
have calculated a different ratio, $R^{part}_{AA}$, defined similarly 
to $R_{AA}$ but with $N_{coll}$ replaced by the number of participant 
pairs, $N_{part}/2$. If particle production increases proportional to the 
number of participants, $R^{part}_{AA} = 1$. 

The obtained $R^{part}_{AA}$ values are shown in Fig.~\ref{fig:yields} 
(bottom) for the three $p_T$ bins used above. The values of
$R_{AA}^{part}$ are larger than $R_{AA}$ by a factor equal to 
$2N_{coll}/N_{part}$ the average number of
nucleon-nucleon collisions suffered by each participant. For 
all $p_T$ bins the yield per participant is consistent with
unity for peripheral collisions as expected since peripheral collisions
should closely resemble N+N collisions. For central collisions 
$R^{part}_{AA}$ increases to approximately three. Most of 
this change occurs in the range of $N_{part}$ from 40 to 140. 
For larger $N_{part}$ the yield in the highest 
$p_T$ bin is approximately constant while in both lower $p_T$ bins it 
increases by 20 to 30\%.

%
%
%
\section{Concluding discussion}
In this paper we have presented the centrality dependence of charged 
hadron $p_T$ spectra focusing on the behavior 
of the spectra at high $p_T$. A striking change of the spectral shape 
is observed when comparing spectra from different centrality 
selections. For peripheral collisions the spectrum exhibits a pronounced 
concave shape which is modified towards a more exponential spectrum as 
the centrality increases. 
The observed lack of variation with centrality in the proton to charged
ratio at large transverse momenta
indicates that the modification is not due to a change in the relative yields
of protons.

%
%
We observe a decrease of $\langle p_T^{trunc} \rangle$ for 
$p_T > 2$~GeV/c with increasing centrality, which is distinctly 
different from the increase of $\langle p_T \rangle$ and demonstrates
the suppression of the high $p_T$ hadron yield independent of
systematic errors associated with the absolute normalization of the spectra. 
The data are not consistent with binary collision scaling of hard scattering 
processes, which would results in an increase of  $\langle p_T^{trunc}
\rangle$. If the $p_T$ spectra above 2~GeV/c are strongly affected by 
collective motion of matter before freeze-out, we would also expect an 
increase $\langle p_T^{trunc} \rangle$ since the corresponding flow 
velocities should increase in more central collisions \cite{edward}.
Similarly, if gluon saturation is important for particle production in 
the $p_T$ range above 2 GeV/c, $\langle p_T^{trunc} \rangle$ should 
increase with increasing $N_{part}$ due to the predicted logarithmic
increase of $Q_s$ \cite{dima2}. In contrast, the data are consistent
with models assuming energy loss of hard scattered partons, which
results in an increasing reduction of the hard scattering contribution 
to the hadron spectrum with increasing centrality of the collisions
\cite{wang-2,GLV,GLV-discovery}. It remains to be seen whether this 
explanation is unique.

%
%
Comparing the measured differential yields in five centrality bins to 
an $N_{coll}$ scaling of the N+N reference yields we see a suppression 
of the yields in central collisions at high $p_T$, consistent with the 
results in \cite{ppg003,star}. In the 0-30\% centrality range (bins 1-3) 
the suppression is approximately independent of $p_T$ for 
$2 < p_T < 5$~GeV/c at a value of $R_{AA} \sim$ 0.6 and simultaneously
nearly independent within 20\% of centrality. The suppression sets in
gradually with the largest change occurring over the 30-60\% centrality 
range. This centrality bin covers a broad range of collision 
geometries. Whether the change is continuous or exhibits a threshold 
behavior, as predicted in \cite{centrality-dependence}, cannot be
judged from the present data. 
The observed suppression is consistent with parton energy loss
scenarios. In these models, the value of $R_{AA}$ and its $p_T$ 
dependence are very sensitive to the actual energy loss prescription. 
Due to the large systematic errors on the $R_{AA}$ scale,
the contribution from the protons and the limited $p_T$ reach of the data 
presented here, we can not distinguish between the different energy
loss prescriptions on the basis of $R_{AA}$.

In summary, a detailed analysis of the centrality dependence of 
charged particle data from Au-Au collisions at $\sqrt{s_{NN}}=130$ GeV
measured by PHENIX reveals interesting features of the 
observed high $p_T$ hadron suppression. The decrease of the 
average $p_T$ with increasing centrality seems 
to favor models of particle production that consider energy 
loss effects, rather than saturation- or hydrodynamics-based 
approaches for this $p_T$ range. The suppression sets in gradually
with the largest changes occurring for peripheral collisions with 
less than about 140 participating
nucleons. From there on it does not change substantially towards 
more central collisions. 
 
%

\begin{ack}
    We thank the staff of the Collider-Accelerator and Physics Departments at
    BNL for their vital contributions.  We acknowledge support from the
    Department of Energy and NSF (U.S.A.), MEXT and JSPS (Japan), RAS,
    RMAE, and RMS (Russia), BMBF, DAAD, and AvH (Germany), VR and KAW
    (Sweden), MIST and NSERC (Canada), CNPq and FAPESP (Brazil), IN2P3/CNRS
    (France), DAE and DST (India), KRF and CHEP (Korea), the U.S. CRDF for
    the FSU, and the US-Israel BSF.
\end{ack}

{\bf Note Added in Proof}

\begin{figure}
  \begin{center}
      \epsfig{file=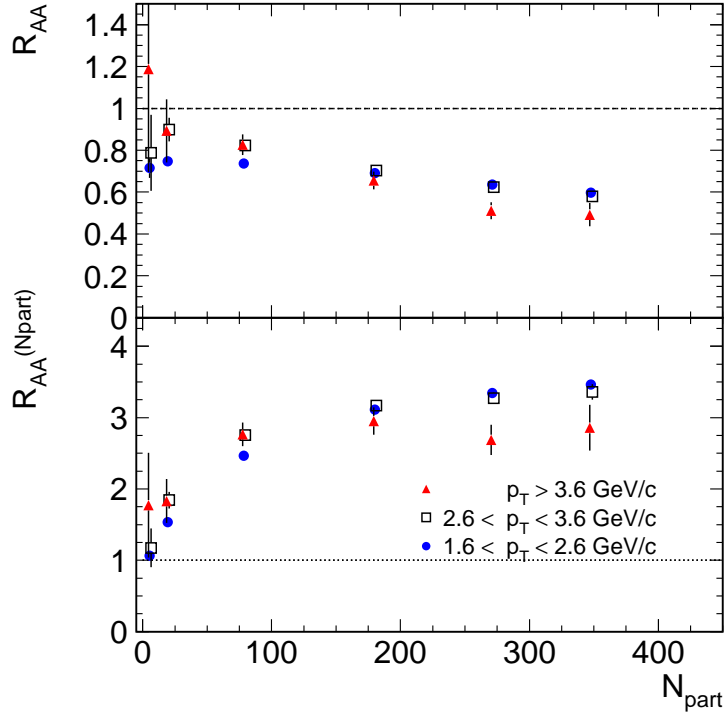,width=0.8\linewidth}
      \caption{\label{fig:yields}
	The top panel gives the nuclear modification factor  $R_{AA}$ 
	for three exclusive $p_T$ regions as a function of the centrality 
	of the collision. The lower panel shows essentially the same quantity 
	but normalized to the number of participant pairs rather than 
	to the number of binary collisions. The dotted line indicates the 
	expectation for scaling with the number of binary collisions
        (top) or with the number of participants (bottom). Only
        statistical errors are shown. The systematic
        error on the scale and the centrality dependence are identical
        to the  errors shown in Fig.~\ref{fig:raa}. These errors are 
        correlated, i.e. take their maximum or minimum value
        simultaneously for all centrality and $p_T$ selections. In
        addition, there are also $p_T$ dependent systematic errors, 
        which are given in Table~\ref{tab:sys}. The systematic errors
        do not alter the trends in the data. 
	}
  \end{center}
\end{figure}

After submission of our manuscript,  data 
from Au-Au collisions at $\sqrt{s_{NN}}=200$~GeV were presented
by the PHOBOS collaboration\cite{Roland:2002un,Back:2003qr}, 
with an emphasis on the lack of variation in the scaled yields with
$N_{part}$ for $N_{part}> 65$. Keeping in mind that results from different
energies are not directly comparable, we note that the broader range in
$N_{part}$ presented here shows that this effect does not apply over the
entire range of centralities, and that when normalized with the
appropriate p-p yields, is simply an aspect of the smooth variation of
$R^{part}_{AA}$ visible in Fig.~\ref{fig:yields}.
 


\vspace{-1cm}

\vspace{-1cm}

\begin{table}
\begin{center} 
\begin{tabular}[]{cccccc}
$p_T$ (GeV/c) & $\delta_{track}$ (\%) & $\delta_{decay}$ (\%) & 
		$\delta_{reso}$ (\%)  & $\delta_{bgr}$ (\%)   &
		total\\ \hline 
1 & $\pm$13.5 & + 10   & $\pm$ 0  & 0     & -13.5 +16.4 \\ 
2 & $\pm$13.5 & +  5   & $\pm$ 1  & 0     & -13.7 +14.4 \\ 
3 & $\pm$13.5 & + 2.5  & $\pm$ 4  & -1.6  & -14.2 +14.2 \\ 
4 & $\pm$13.5 & + 1.25 & $\pm$ 9  & -11.5 & -20   +16   \\ 
5 & $\pm$13.5 & + 0.6  & $\pm$ 15 & -40   & -45   +20   \\ 
\end{tabular}
\caption{ \label{tab:sys} Upper bounds of the
systematic error on the $p_T$ dependent single particle correction
function. Here $\delta_{track}$  
includes the uncertainties of the acceptance, dead areas, track 
matching cuts and the track reconstruction efficiency. The
$\delta_{decay}$ term accounts for the uncertainty of the decay
correction. The effect of the momentum resolution contributes with
$\delta_{reso}$ to the systematic error. Uncertainties due to
potentially unsubtracted background are quantified by $\delta_{bgr}$. 
The total systematic error given in the last column is calculated as
quadrature sum of the individual contributions. It is calculated
separately for positive and negative errors. 
}
\end{center}
\end{table}

\begin{table}
\begin{center} 
\begin{tabular}[]{cccccc}
bin & relative fraction 
& $N_{part}$ & $N_{coll}$ & $N_{coll}^{central}/N_{coll}$ 
& $2 N_{coll}/N_{part}$ \\ \hline
1& 80-92\% & 5.5$\pm$ 2.6 & 4.1$\pm$1.7  & 246$\pm$98    & 1.5$\pm$0.5 \\ 
2& 60-80\% & 19.5$\pm$ 3.5 &  20$\pm$ 6  & 50.4$\pm$13   & 2.1$\pm$0.5 \\ 
3& 30-60\% &  79$\pm$ 4.6 & 131$\pm$23   & 7.68$\pm$1.1  & 3.4$\pm$0.6 \\ 
4& 15-30\% & 180$\pm$ 6.6 & 406$\pm$46   & 2.49$\pm$0.13 & 4.5$\pm$0.5 \\ 
5& 5-15\%  & 271$\pm$ 9   & 712$\pm$72   & 1.41$\pm$0.03 & 5.2$\pm$0.6 \\ 
6& 0-5\%   & 348$\pm$10   & 1009$\pm$101 & 1             & 5.8$\pm$0.6 \\	
\end{tabular}
\caption{ \label{tab:1} Number of participants and binary collisions 
and their systematic errors for the individual centrality selections used 
in this analysis. Also given is the ratio of the number of binary collisions
for the most central sample relative to the one for each sample. 
The last column 
quantifies the ratio of binary collisions to participant pairs.}

\end{center}
\end{table}


\begin{thebibliography}{10}

\def\IJMPA{{Int. J. Mod. Phys.}~{\bf A}}
\def\JPG{{J. Phys}~{\bf G}}
\def\NCA{Nuovo Cimento}
\def\NIM{Nucl. Instrum. Methods}
\def\NIMA{{Nucl. Instrum. Methods}~{\bf A}}
\def\NPA{{Nucl. Phys.}~{\bf A}}
\def\NPB{{Nucl. Phys.}~{\bf B}}
\def\PLB{{Phys. Lett.}~{\bf B}}
\def\PLC{Phys. Repts.\ }
\def\PRL{Phys. Rev. Lett.\ }
\def\PRD{{Phys. Rev.}~{\bf D}}
\def\PRC{{Phys. Rev.}~{\bf C}}
\def\ZPC{{Z. Phys.}~{\bf C}}
\def\EPJ{Eur.Phys.J.~{\bf C}}

\bibitem[*]{Deceased}Deceased     
\bibitem[$^{\dagger}$]{nonpart}Not a participating Institution.  

\bibitem{E605}          H.J{\"o}stlein et al., \PRD{20} (1979) 53.

\bibitem{jet-quenching} M.~Gyulassy and M.~Pl\"umer, \PLB{243} (1990) 432;
			X.N.~Wang and M.~Gyulassy, \PRL{68} (1992) 1480.
\bibitem{quench-density-dep} X.N.~Wang, M.~Gyulassy and M.~Pl\"umer, 
				\PRD{51} (1995) 3436; 
			R.~Baier et al., \PLB{345} (1995) 277; 
			R.~Baier, D.~Schiff and B.G.~Zakharov, 
			Ann.~Rev.~Nucl.~Part.~Sci.~{\bf 50} (2000) 37-69.
\bibitem{quench-effect} M.~Gyulassy and X.N.~Wang, \NPB{420} (1994) 583;
			X.N.~Wang, \PRC{58} (1998) 2321.
\bibitem{ppg003}	PHENIX Collaboration: K.~Adcox  et al., 
			\PRL{88} (2002) 22301.
\bibitem{star}		STAR Collaboration: C.Adler et al.,
			\PRL{89} (2002) 202301.
\bibitem{wang-2}        X.N.~Wang \PRC{61} (2000) 64910.
\bibitem{GLV}		M.~Gyulassy, P.~Levai and I.~Vitev, 
			\PRL{85} (2000) 5535.
\bibitem{GLV-discovery} P.~Levai et al., \NPA{698} (2002) 631. 
\bibitem{gallmeister}	K.~Gallmeister, C.~Greiner, and Z.~Xu,
			nucl-th/0202051.
\bibitem{kolb}	        P.~Kolb et al., \NPA{696} (2001) 197. 
\bibitem{teaney}        D.~Teaney, J.~Lauret and E.V.~Shuryak, nucl-th/0110037.
\bibitem{EMC}		J.J.~Aubert et al., \PLB{123}, (1983) 275.
\bibitem{Gribov}	L.V.~Gribov, E.M.~Levin and M.G.~Ryskin,
			\PLC{100} (1983) .
\bibitem{sarcevic}      Z.~Huang, H.~J.~Lu and I.~Sarcevic,
			\NPA{637} (1998) 79.
\bibitem{eskola}	K.J.~Eskola, V.J.~Kolhinen and C.A.~Salgado, 
			\EPJ{9} (1999) 61.

\bibitem{mclerran}      L.~D.~McLerran and R.~Venugopalan,
			\PRD{49} (1994) 2233.
\bibitem{kovner}        A.~Kovner, L.~D.~McLerran and H.~Weigert,
			\PRD{52} (1995) 6231.
%
\bibitem{krasnitz}      A.~Krasnitz, Y.~Nara and R.~Venugopalan,
			\PRL{87} (2001) 192302.
\bibitem{mclerran-2}	D.~Kharzeev, E.~Levin and L.~D.~McLerran,
			hep-ph/0210332.
%
\bibitem{ppg002}        PHENIX Collaboration: K.~Adcox et al., 
			\PRL{86} (2001) 3500 and \PRL{87} (2001) 052301.
\bibitem{ppg006}        PHENIX Collaboration: K.~Adcox et al.,  
			\PRL{88} (2002) 242301.
\bibitem{wang-enke} 	E.~Wang and X.~N.~Wang, nucl-th/0104031 and
			references therein.
\bibitem{cronin}        D.~Antreasyan {\it et al.}, \PRD{19} (1979) 764.
\bibitem{edward}	 E.~Shuryak private communication. 
\bibitem{dima2}		D.~Kharzeev and E.~Levin, nucl-th/0108006, 
			J.~Schaffner-Bielich et al., nucl-th/0108048.
\bibitem{centrality-dependence} X.N.~Wang, \PRC{63} (2001) 54902. 

\bibitem{Roland:2002un}  PHOBOS Collaboration: C.~Roland et al.,  
arXiv:hep-ex/0212006.

\bibitem{Back:2003qr}    PHOBOS Collaboration: B.~B.~Back et al., 
arXiv:nucl-ex/0302015, submitted to Phys.~Lett.~B.

\end{thebibliography}
\end{document}